\documentclass[a4paper,11pt]{article}
\usepackage[dvips]{graphicx}
\usepackage{amssymb}
\usepackage{amsmath}
\usepackage{cite}

\begin{document}

\title{Superbounce and Loop Quantum Cosmology Ekpyrosis from Modified Gravity}
\author{
V.K. Oikonomou$^{1}$\,\thanks{v.k.oikonomou1979@gmail.com}\\
$^{1)}$Department of Theoretical Physics, Aristotle University of Thessaloniki,\\
54124 Thessaloniki, Greece
} \maketitle

\begin{abstract}
As is known, in modified cosmological theories of gravity many of the cosmologies which could not be generated by standard Einstein gravity, can be consistently described by $F(R)$ theories. Using known reconstruction techniques, we investigate which $F(R)$ theories can lead to a Hubble parameter describing two types of cosmological bounces, the superbounce model, related to supergravity and non-supersymmetric models of contracting ekpyrosis and also the Loop Quantum Cosmology modified ekpyrotic model. Since our method is an approximate method, we investigate the problem at large and small curvatures. As we evince, both models yield power law reconstructed $F(R)$ gravities, with the most interesting new feature being that both lead to accelerating cosmologies, in the large curvature approximation. The mathematical properties of the some Friedmann-Robertson-Walker spacetimes $M$, that describe superbounce-like cosmologies are also pointed out, with regards to the group of curvature collineations $CC(M)$.  
\end{abstract}


\section*{Introduction}

Having to deal with singularities is one of the most difficult task in all physical systems that these occur, and the theoretical explanation of the appearance of these can be quite hard, or the window to new physics. Cosmological singularities occur in many theoretical frameworks and a question rises, whether these are a real ingredient of the cosmological expansion or these are an indicator that the classical theory needs to be implemented with quantum gravity effects, in order these are treated properly, if not completely disappear. So in modern cosmology there appeared two types of non-singular descriptions of the cosmological expansion, the ones that address the problem by using the Loop Quantum Cosmology (abbreviated to LQC hereafter) framework \cite{LQC1,LQC2,LQC3}, thus residing to a quantum gravitational description of the cosmos, and the other models that use scalar fields to avoid singular expansion \cite{bounce1,bounce2,bounce3,bounce4,bounce5,ekpyr1,ekpyr2,ekpyr3}. With respect to LQC, singularities are amended by using Loop Quantum Gravity techniques, such as holonomy corrections to existing cosmological models. For recent approaches to cosmological aspects of quantum gravity see \cite{oriti}. In both cases, the resulting expansion is a bounce with expanding and contracting phases, with the bounce providing an alternating description to the acceleration and thermal history of our universe. For recent informative reviews on cosmological bounces, the reader is referred to \cite{bounce1,bounce2} and references therein. In addition, the bounce models are known to give a consistent explanation of certain CMB anomalies on large angular scale. For an interesting account on this, see \cite{piao}.

In general, bounce models carry along some problems which have to be resolved in order a more consistent behavior is achieved. For example, in order a bounce occurs in a scalar field modified gravity context, the null energy condition has to be violated for all the matter fields considered in the theoretical framework, and particularly the sum of the pressure and matter energy density has to be negative. Notice that this is possible in systems for which the Hamiltonian is bounded from below, see \cite{vikman2}. If the null energy condition is violated, then it is possible that the Hubble parameter becomes negative and a contracting phase is possible for the cosmological evolution \cite{bounce1,bounce2,bounce4,bounce5}. In addition, the bounce models suffer from ghost instabilities and from primordial instabilities which occur during the contracting phase, with the latter being a somewhat problematic phase of the cosmological expansion. Ghost instabilities are resolved \cite{vikman1} in the Galileon and ghost condensate models \cite{vikman1,bounce3} but the contracting phase still remains an obstacle to providing a fully consistent description of the cosmos expansion. To this end, in references \cite{bounce4,bounce5} an ekpyrosis scenario for the contracting phase was adopted, in which problems related with specific initial conditions are avoided. In reference \cite{bounce4} a model of contracting ekpyrosis was presented in both supergravity and non-supersymmetric frameworks and in \cite{bounce5} the cosmological perturbations of the contracting ekpyrosis model was addressed, among other issues.

In view of the cosmological appealing solution that the contracting ekpyrosis provides, in this paper we shall investigate which $F(R)$ gravity can generate such a cosmological expansion. Particularly, we shall focus our search on which $F(R)$ gravity generates the Hubble parameter that corresponds to the contracting ekpyrosis scenario of references \cite{bounce4,bounce5}. The method we shall use is a reconstruction method quite well known in the literature \cite{importantpapers3} and applying this quite general technique, we shall study the large and small curvature limits of the $F(R)$ gravity that generates the Hubble expansion of the contracting ekpyrosis. Since this is an approximate method, the exact scale factor of the contracting ekpyrosis in the classical limits of $t\gg t_0$ and $t\ll t_0$ can be generated by the corresponding $F(R)$ gravities, with $t_0$ the time instance at which the bounce occurs.

Many cosmological scenarios which was impossible to be realized in classical Einstein gravity, can be consistently realized in the context of $F(R)$ gravities. For an important stream of review articles and important papers on $F(R)$ theories, see \cite{reviews1,importantpapers1,importantpapers2,importantpapers3,importantpapers4,recontechniques,sergeinojirimodel,sergeibabmba} and references therein. Among various elegant descriptions of large scale gravitational effects that the $F(R)$ models describe, one of the most important successes is the description of dark energy and inflation in the same theoretical framework, with first model materializing this being the Nojiri-Odintsov model \cite{sergeinojirimodel}. Note that there is interesting consequence of result we discover in this paper.
As we will see, the key point in our study, in the case of the superbounce, will be played by $R^2$ terms in the $F(R)$ gravity and at the same time, as was shown in \cite{sergeinojirimodel}, an $R^2$ term models precisely the unification of inflation with dark energy. Furthermore, it has been proven that $R^2$ term not only causes the appearance of inflationary sector but also removes finite-time future singularity \cite{sergeibabmba}. In addition, power law corrections are known to link certain limits of chameleon cosmology \cite{chameleon} with $F(R)$ gravity.

For alternative to $F(R)$ theories, that also consistently address the dark energy problem, see \cite{capo,capo1,peebles,faraonquin,tsujiintjd,wetterich}. In addition, in the cosmon-related theories constructed by Wetterich, inflation and dark energy are also described by the same theory \cite{wetterich}.

The theoretical challenges raised in modern theoretical cosmology after the late 90's striking observation of the universe's late time expansion \cite{riess}, have generated increasing interest the last decade and with the aid of current observational searches, many of these theories are going to critically tested. Particularly, among other issues, inflation, dark energy and the bounce models can be tested by the BICEP \cite{bicep} and Planck \cite{planck} observational projects. With respect to the bounce models, there are promising and literally down to earth proposed experiments \cite{kinezosvergados} that may reveal the bounce using direct dark matter scattering on nuclei. For an account on the latter issue see \cite{oikonomouvergados}. Therefore, it is of great importance to have many alternative theoretical models for the same cosmological evolution of the universe. This is our motivation for this study, and in this article we shall try to generate the superbounce ekpyrosis from $F(R)$ gravity. In addition, we also reconstruct the $F(R)$ gravity that generates the ekpyrotic scenario that results from a LQC framework.  It is quite intriguing that in both cases, the large curvature limit of the reconstructed $F(R)$ gravity leads to accelerating cosmologies.


This paper is organized as follows: In section 1, we reconstruct the $F(R)$ gravity that generates the Hubble parameter corresponding to the superbounce model developed in \cite{bounce4} in the large and small curvature limits. We also examine the resulting $F(R)$ on how it is possible to generate from it, early time and late time acceleration in the Jordan frame. A special mathematical property of a specific spacetime that belongs to the same class of Friedmann-Robertson-Walker metric (FRW hereafter) spacetimes that the superbounce belongs is also studied. In section 2, we use the same reconstruction technique in order to find which $F(R)$ gravity can generate the LQC ekpyrotic cosmological scenario. The conclusions follows in the end of the paper.

\subsubsection*{Conventions}

Throughout the article, if it is not mentioned, we shall make use of the following spacetime geometry conventions. Particularly, we shall assume that the $F(R)$ gravities are studied in the Jordan frame and we shall make use of the metric formalism in order to obtain the equations of motion \cite{reviews1}. Moreover, it shall be assumed that the geometric background is that of a pseudo-Riemannian manifold, locally being described by a Lorentz metric, a flat FRW, which is,
\begin{equation}\label{metricformfrwhjkh}
\mathrm{d}s^2=-\mathrm{d}t^2+a^2(t)\sum_i\mathrm{d}x_i^2
\end{equation}
The Ricci scalar corresponding to metric (\ref{metricformfrwhjkh}) is,
\begin{equation}\label{ricciscal}
R=6(2H^2+\dot{H}),
\end{equation}
with $H(t)$ being the Hubble parameter and the ``dot'' denoting as usual, differentiation with respect to the cosmic time $t$. Moreover, the affine connection of the manifold is assumed to be the Levi-Civita connection, which is torsion-less, symmetric, and
metric compatible.

\section{Superbounce from $F(R)$ Gravity-Small and Large Curvature Limits}

\subsection{The non-Supersymmetric Super-bounce Scenario}

The non supersymmetric superbounce solution, results from a bounce Lagrangian of the form \cite{bounce4,bounce5},
\begin{equation}\label{superblagra}
\mathcal{L}=\sqrt{-g}\left ( -\frac{R}{2}+P(X,\phi)+g(\phi )X\square \phi \right )
\end{equation} 
with $R$ the Ricci scalar, $X=-\frac{1}{2}(\partial \phi )^2$ and $P(X,\phi)$ being equal to,
\begin{equation}\label{spfx}
P(X,\phi)=k(\phi)X+\tau(\phi)X^2-V(\phi)
\end{equation}
The function $k(\phi )$ is chosen in a way such that it is almost everywhere equal to one except around $\phi=0$, where the bounce occurs and is assumed that it takes the following form \cite{bounce4},
\begin{equation}\label{kphi}
k(\phi )=1-\frac{2}{(1+2k\phi^2)^2}
\end{equation}
with the parameter $k$ controlling the width in the $\phi$-space, for which the field kinetic term does not change sign. In addition, the function $\tau(\phi)$ in relation (\ref{spfx}), controls the strength of the square of the kinetic term and $g(\phi )$ controls the strength of the Galileon term. These are chosen in the following way \cite{bounce4},
\begin{equation}\label{supergtau}
\tau(\phi )\sim \frac{\bar{\tau}}{(1+2k\phi^2)^2},{\,}{\,}{\,}g(\phi )= \frac{\bar{g}}{(1+2k\phi^2)^2}
\end{equation}
Finally, the potential $V(\phi)$ in relation (\ref{spfx}) is chosen for large values of the field parameter $\phi$ to be,
\begin{equation}\label{largephi}
V(\phi)=-V_0v(\phi)e^{-c(\phi)\phi}
\end{equation}
with $v(\phi)$ a function chosen in such a way so that the potential turns off for $\phi$ approaching the ending ekpyrotic value. The evolution of the universe in such a background is given by,
\begin{equation}\label{basicsol1}
a(t)\sim (-t+t_*)^{2/c^2}
\end{equation}
with $t_*$ denoting the big crunch time, if the ekpyrotic phase continues until that time. The solution (\ref{basicsol1}) describes the background evolution for large $\phi$ values. At values of $\phi$ near the ekpyrotic bounce ending, the background solution reads,
\begin{equation}\label{basicsol12}
a(t)\sim (-t+t_*)^{1/3}
\end{equation}
The Hubble parameter corresponding to the ekpyrotic scaling solution (\ref{basicsol1}) equals to,
\begin{equation}\label{hubble1}
H(t)=\frac{2}{c^2(t-t_*)}
\end{equation}
while the one corresponding to (\ref{basicsol12}) is equal to,
\begin{equation}\label{hubble12}
H(t)=\frac{1}{3(t-t_*)}
\end{equation}
It is the main focus of this section to construct $F(R)$ gravities which produce the cosmology given from relations (\ref{hubble1}) and (\ref{hubble12}). We mainly study the case described by relations (\ref{basicsol1}) and (\ref{hubble1}) since the other case can be easily obtained from the one we study.

Before proceeding, we have to note that, as was demonstrated in \cite{bounce4}, the supergravity extension of (\ref{largephi}) exactly reproduces the cosmology given in (\ref{basicsol1}) and this justifies the terminology superbounce that we used for it.

\subsection{Reconstruction of $F(R)$ Gravity}

From both relations (\ref{hubble1}) and (\ref{hubble12}) we can easily elicit the bounce behavior and it worths for a moment to clarify the values that time can have. Practically speaking, $t$ varies from the time $-t_*$ to $+t_*$, as can be seen from the behavior of (\ref{hubble1}). In addition, the same can be deduced by looking the scale factor (\ref{basicsol1}). Now it is of critical importance to settle what we mean by large and small time limits $t$. With large $t$ we mean for large cosmic times before the crunch and small times refer to times just after the beginning of the bounce. Of course it is not our purpose to describe both scale solutions (\ref{basicsol1}) and (\ref{basicsol12}) with the same $F(R)$ gravity, since that would probably require varying coefficients probably for the $F(R)$ function as we shall see. So the full ekpyrotic superbounce solution is much too involved, but here we are just interested for it's limiting behavior. This is an important notice since in the context of the ekpyrotic superbounce scenario, the universe's cosmic time might not even reach the crunch time $t_*$. 

It is the purpose of this section to reconstruct the $F(R)$ gravities that produce the cosmology given by the Hubble parameter (\ref{hubble1}) in the large and small cosmic time limits. We shall use the reconstruction method firstly developed in \cite{importantpapers3}, which heavily relies on a particular form that the hubble parameter takes. For other important reconstruction techniques see \cite{recontechniques}. The technique developed in \cite{importantpapers3} is also applied whenever the Hubble parameter can be written as $\sim \frac{h(t)}{t}$, with $h(t)$ a slowly varying function of time. The fact that $h(t)$ is considered to be a slowly varying function of time, makes possible some sort of perturbation at the level of Euler-Lagrange equations. In the case at hand, the Hubble parameter (\ref{hubble1}) can be written as follows,
\begin{equation}\label{hubble1two}
H(t)=\frac{2}{c^2(t-t_*)}=\frac{2qt}{c^2t(qt-qt_*)}=\frac{h_fqt}{t(qt-qt_*)}=\frac{h(t)}{t}
\end{equation}
with $h_f=2/c^2$ and $h(t)$ being equal to,
\begin{equation}\label{ht}
h(t)=\frac{h_fqt}{(qt-b)}
\end{equation}
with $b=qt_*$ and $q$ is assumed to be very small, that is $q\ll 1$. The function $h(t)$ is a slowly varying function of $t$ and therefore the reconstruction method of reference \cite{importantpapers3} applies. Indeed, $h(t)$ satisfies $\forall $ $z$ $\in $ $\mathbb{R}$, the following relation,
\begin{equation}\label{r10}
\lim_{t\rightarrow \infty}\frac{h(zt)}{h(t)}=1
\end{equation}
and so it is a slowly varying function. Let us introduce the theoretical framework of the reconstruction method and then we apply it straightforwardly. Assume that the action of the general $F(R)$ gravity is,
\begin{equation}\label{action1dse}
\mathcal{S}=\frac{1}{2\kappa^2}\int \mathrm{d}^4x\sqrt{-g}F(R)+S_m(g_{\mu \nu},\Psi_m),
\end{equation}
while the FRW equation is:
\begin{equation}\label{frwf1}
-18\left ( 4H(t)^2\dot{H}(t)+H(t)\ddot{H}(t)\right )F''(R)+3\left (H^2(t)+\dot{H}(t) \right )F'(R)-\frac{F(R)}{2}+\kappa^2\rho=0
\end{equation}
with $F'(R)=\frac{\mathrm{d}F(R)}{\mathrm{d}R}$ and the Ricci scalar $R$ is given in (\ref{ricciscal}). We assume in addition that the mass-energy density receives contribution from all kind of perfect matter fluids, with cosmological parameters $w_i$. The reconstruction method of reference \cite{importantpapers3} makes use of an auxiliary scalar field $\phi $, which is introduced directly into action (\ref{action1dse}), which becomes,
\begin{equation}\label{neweqn123}
S=\int \mathrm{d}^4x\sqrt{-g}\left ( P(\phi )R+Q(\phi ) +\mathcal{L}_{mat} \right )
\end{equation}
The final form of the reconstructed $F(R)$ gravity shall be given by the functions $P(\phi )$ and $Q(\phi )$, so finding them is the main goal. The lack of a kinetic term render the scalar field $\phi$ an auxiliary degree of freedom, and upon variation with respect to $\phi$, action (\ref{neweqn123}) reads,
\begin{equation}\label{auxiliaryeqns}
P'(\phi )R+Q'(\phi )=0
\end{equation}
where the prime denotes differentiation with respect to $\phi$ in this case. By solving equation (\ref{auxiliaryeqns}) with respect to $R$, we obtain $\phi (R)$ and by substituting it to the original $F(R)$ action, we get the $F(R)$, that is,
\begin{equation}\label{r1}
F(\phi( R))= P (\phi (R))R+Q (\phi (R))
\end{equation}
Practically speaking, by finding a differential equation for $P(\phi )$ and $Q(\phi )$, will yield the $F(R)$ gravity, given the Hubble parameter and a specific form of the scale factor. This point is crucial in our analysis, in reference to the specific form of the scale factor, and we shall discuss it further, later on in this section. By varying equation (\ref{neweqn123}) with respect to the metric tensor and assuming a spatially flat FRW metric, we obtain the following differential equation for $P(\phi )$ and $Q(\phi )$,
\begin{align}\label{r2}
& -6H^2P(\phi (t))-Q(\phi (t) )-6H\frac{\mathrm{d}P\left (\phi (t)\right )}{\mathrm{d}t}+\rho_i=0 \\ \notag &
\left ( 4\dot{H}+6H^2 \right ) P(\phi (t))+Q(\phi (t) )+2\frac{\mathrm{d}^2P(\phi (t))}{\mathrm {d}t^2}+\frac{\mathrm{d}P(\phi (t))}{\mathrm{d}t}+p_i=0
\end{align}
and by eliminating $Q(\phi (t))$ we obtain,
\begin{equation}\label{r3}
2\frac{\mathrm{d}^2P(\phi (t))}{\mathrm {d}t^2}-2H(t) P(\phi (t))+4\dot{H}\frac{\mathrm{d}P(\phi (t))}{\mathrm{d}t}+\rho_i+p_i=0
\end{equation}
where $\rho_i,p_i$ stand for the mass-energy density and pressure of the matter perfect fluids. As was proven in reference \cite{importantpapers3}, owing to the equivalence of the actions (\ref{action1dse}) and (\ref{neweqn123}), the scalar field $\phi$ can be identified with the cosmic time $t$, that is, $\phi=t$ (see \cite{importantpapers3} for details). In the following we shall switch between the two variables which we consider one and the same. Now a crucial point, we shall assume a generic behavior for the scale factor, being of the form,
\begin{equation}\label{r4}
a=a_0e^{g(t)}
\end{equation}
with $a_0$ constant. The differential equation (\ref{r3}) for $P(\phi )$ is therefore written as follows,
\begin{align}\label{r5}
& 2\frac{\mathrm{d}^2P(\phi (t))}{\mathrm {d}t^2}-2g'(\phi )\frac{\mathrm{d}P(\phi (t))}{\mathrm{d}t}+4g''(\phi ) P(\phi (t))+\sum_i(1+w_i)\rho_{i0}a_0^{-3(1+w_i)}e^{-3(1+w_i)g(\phi )}=0
\end{align}
We have to mention that the superbounce scale factor can be identical to the form (\ref{r4}) only in the large $t$ and small $t$ limits, as can be checked, but our interest is to reproduce the cosmology described by the Hubble parameter corresponding to the superbounce.  

Having found $P(\phi )$ by solving the differential equation (\ref{r5}), it is easy to obtain $Q(\phi )$, by using relation (\ref{r2}), and the resulting relation is,
\begin{equation}\label{r5a}
Q(\phi )=-6g'(\phi )^2P(\phi )-6g'(\phi )\frac{\mathrm{d}P(\phi )}{\mathrm{d}\phi }+\sum_i(1+w_i)\rho_{i0}a_0^{-3(1+w_i)}e^{-3(1+w_i)g(\phi )}
\end{equation}
Now by exploiting the fact that the Hubble parameter is written as a fraction of a slowly varying function of time over time, given in relation (\ref{hubble1two}), we assume that the function $g(t)$ appearing in (\ref{r4}), takes the following form,
\begin{equation}\label{r11}
g(\phi )=h(\phi)\ln \left( \frac{\phi}{\phi_0}\right )
\end{equation}
with $\phi_0$ some constant. A consequence of the fact that $h(t)$ is slowly varying is that we can neglect the higher derivatives $h'(t)$, $h''(t)$, etc. This means that the Hubble parameter corresponding to the scale factor given in (\ref{r4}), with  $g(t)$ given in (\ref{r11}), can take the following form,
\begin{equation}\label{r11a}
H(t)=\frac{h(t)}{t}+h'(t)\ln\left (\frac{t}{t_0} \right ) 
\end{equation} 
which, by neglecting the derivative can be written,
\begin{equation}\label{r11aguns}
H(t)=\frac{h(t)}{t} 
\end{equation} 
This is a crucial point in our analysis, because owing to the fact that the superbounce Hubble parameter can be written exactly in the form of relation (\ref{r11aguns}) (without any approximation) enables us to use the present reconstruction technique and this is what motivated us to apply this method. Returning to the differential equation (\ref{r11}), by substituting $g(t)$ and ignoring the higher derivatives $h'(t),h''(t)$, we obtain, 
\begin{align}\label{r12}
& 2\frac{\mathrm{d}^2P(\phi (t))}{\mathrm {d}t^2}-\frac{h(\phi )}{\phi }\frac{\mathrm{d}P(\phi (t))}{\mathrm{d}t}-\frac{2h(\phi )}{\phi^2} P(\phi (t))
\\ \notag & +\sum_i(1+w_i)\rho_{i0}a_0^{-3(1+w_i)}\left ( \frac{\phi }{\phi_0}\right )^{-3(1+w_i)h(\phi )}=0
\end{align} 
The key point of the whole calculation is to express the Ricci scalar as a function of $\phi$ (or equivalently $t$) and invert the resulting expression so that we have at hand $\phi=\phi(R)$. After that it is easy to find the functions $P(\phi (R))$ and $Q(\phi (R))$ and hence, the $F(R)$ gravity that generates the specific Hubble expansion. Using relations (\ref{ricciscal}), (\ref{hubble1two}) and (\ref{r11}), and by ignoring the higher order derivatives for the slowly varying function $h(t)$, the Ricci scalar, which for a FRW is given in relation (\ref{ricciscal}), is now equal to,
\begin{equation}\label{r13}
R(\phi )\simeq \frac{6\left (-h(\phi )+2h(\phi )^2  \right )}{\phi^2}
\end{equation}
By solving the above equation with respect to $\phi$, will yield $\phi=\phi(R)$. By substituting $h(t)$ from relation, (\ref{ht}), we get the following equation,
\begin{equation}\label{r14}
-b^2 R t^2+2 b q R t^3-q^2 R t^4+6 b q t h_f-6 q^2 t^2 h_f+12 q^2 t^2 h_f^2=0 
\end{equation}
The real solution to this equation is,
\begin{align}\label{r15}
& t = \frac{2b}{q}-\frac{2^{1/3} b^2 R}{3\left ( \mathcal{P}_1+\sqrt{\mathcal{P}_2}\right )^{1/3}}+\frac{6\times 2^{1/3} q^2 h_f}{\left ( \mathcal{P}_1+\sqrt{\mathcal{P}_2}\right )^{1/3}}
-\frac{12\times 2^{1/3} q^2 h_f^2}{\left ( \mathcal{P}_1+\sqrt{\mathcal{P}_2}\right )^{1/3}}-\frac{\left ( \mathcal{P}_1+\sqrt{\mathcal{P}_2}\right )^{1/3}}{3\times 2^(1/3) q^2 R} 
\end{align}
with the polynomials $\mathcal{P}_1$ and $\mathcal{P}_2$ being equal to,
\begin{equation}\label{polynmials}
\mathcal{P}_1=a_1R^2+a_2R^3,{\,}{\,}{\,}\mathcal{P}_2=\beta_3R^3+\beta_4R^4+\beta_5R^5
\end{equation}
The coefficients $\alpha_i,\beta_i$ of the polynomials (\ref{polynmials}) are given in the appendix A. The solution given in relation (\ref{r15}) is very important, since when solving the differential equation (\ref{r12}) we can substitute the obtained solution (\ref{r15}) and find the resulting expressions for $P(\phi(R) )$ and $Q(\phi(R))$. Then by using (\ref{r1}) we have a resulting expression for the reconstructed $F(R)$. The general solution of the differential equation (\ref{r12}) is of the following form \cite{importantpapers3},
\begin{align}\label{r16}
& P(\phi )=c_1\phi^{\frac{h(\phi )-1+\sqrt{h(\phi)^2+6h(\phi )+1}}{2}}+c_2\phi^{\frac{h(\phi )-1-\sqrt{h(\phi)^2+6h(\phi )+1}}{2}}+\sum_iS_i(\phi )\phi^{-3(1+w_i)h(\phi )+2}
\end{align}
with $S_i(\phi )$ being,
\begin{align}\label{r17}
& S_i(\phi )=-\left ((1+w_i)\rho_{i0}a_0^{-3(1+w_i)}\phi_0^{3(1+w)h(\phi )} \right )\\ \notag & \left ( 6(1+w_i)(4+3w_i)h(\phi)^2-2(13+9w)h(\phi)+4\right )^{-1}
\end{align}
Also, from the solution (\ref{r16}), we can find $Q(\phi)$ by simply substituting (\ref{r11}) and (\ref{r17}) in equation (\ref{r5a}). So finally $Q(\phi )$ is equal to,
\begin{align}\label{r18}
& Q(\phi )=-6h(\phi )c_1\left (h(\phi )+\frac{h(\phi )-1+\sqrt{h(\phi)^2+6h(\phi )+1}}{2}\right )\phi^{\frac{h(\phi )-1+\sqrt{h(\phi)^2+6h(\phi )+1}}{2}-2}\\ \notag &
-6h(\phi )c_2\left (h(\phi )+\frac{h(\phi )-1-\sqrt{h(\phi)^2+6h(\phi )+1}}{2}\right )\phi^{\frac{h(\phi )-1+\sqrt{h(\phi)^2+6h(\phi )+1}}{2}-2}\\ \notag & +\sum_i\left ( -6h(\phi )(-2(2+3w)h(\phi )+2+S_i(\phi )+\rho_{i0}a_0^{-3(1+w_i)\phi_0^{3(1+w_i)h(\phi)}})\right )\phi^{-3(1+w_i)h(\phi )}
\end{align}
In principle, finding the general form of the $F(R)$ function can be quite difficult, and therefore it is more convenient to study the problem in the large and small curvature limits which corresponds to small and large $t$ limits respectively. 


 \subsection{$F(R)$ Gravity in the Large $R$ Limit}

We start off with the large curvature $R$ limit of the solutions (\ref{r16}) and (\ref{r18}) we found in the previous section. As already stated, the large curvature limit corresponds to the small $t$ limit, so the solution (\ref{r15}) for $\phi$ in the large curvature limit reads, 
\begin{align}\label{r16a}
& \phi\simeq -\frac{2b}{3 q}+ \frac{2^{1/3}b^{2} R}{3\left (\alpha_2R^3+\sqrt{\beta_5R^5}\right )^{1/3}}-\frac{\left (\alpha_2R^3+\sqrt{\beta_5R^5}\right )^{1/3}}{3\times 2^{1/3}q^2R}
\end{align}
where only the dominant terms of the polynomials $\mathcal{P}_1$ and $\mathcal{P}_2$ are considered. By using the values for the polynomial coefficients $\alpha_2$ and $\beta_5$ given in the appendix A, we can see that the first two terms cancel, and therefore the expression (\ref{r16a}) becomes,
\begin{align}\label{r17b}
& \phi\simeq \frac{\mathcal{A}}{R}
\end{align}
where we have set $\mathcal{A}$ to be equal to,
\begin{align}\label{r18a}
\mathcal{A}=\frac{6\times 2^{1/3} q^2h_f-12\times 2^{1/3}q^2 h_f^2}{3}
\end{align}
Having relation (\ref{r18a}) at hand, we can straightforwardly find the function $P(\phi (R))$, by substituting (\ref{r17b}) in (\ref{r16}). Before doing that, let us simplify further the function $P(\phi )$ by taking the small time (large curvature) limit of relation (\ref{r16}). The small and large $t$ limiting values of the slowly varying function $h(t)$ are,
\begin{equation}\label{r19}
\lim_{t\rightarrow 0}h(t)=0,{\,}{\,}{\,}\lim_{t\rightarrow \infty }h(t)=h_f
\end{equation} 
So, owing to the fact that $\lim_{t\rightarrow \infty}h(t)=0$, the small $t$ limit of $P(\phi )$ is,
\begin{equation}\label{r20}
P(\phi )=c_1+c_2\phi^{-2}+\sum_iS_i(0)\phi^2
\end{equation}
and by using (\ref{r17}), equation (\ref{r20}) becomes,
\begin{equation}\label{r21}
P(\phi (R))=c_1+\frac{c_2}{\mathcal{A}}R+\sum_iS_i(0)\frac{\mathcal{A}^2}{R^2}
\end{equation}
By keeping only the dominant terms in the large curvature limit, $P(\phi (R))$ becomes,
\begin{equation}\label{r22}
P(\phi (R))\simeq c_1+\frac{c_2\alpha_3}{\mathcal{A}_1}R
\end{equation}
In the same vain, the function $Q(\phi (R))$ becomes,
\begin{equation}\label{r23}
Q(\phi(R))=\sum_iS_i(0 )a_0^{-3(1+w_i)}\phi_0^{3(1+w_i)}
\end{equation}
Gathering up the results of relations (\ref{r22}) and (\ref{r23}) and also by setting,
\begin{equation}\label{jfhdgdv}
c_4=\sum_iS_i(0)a_0^{-3(1+w_i)}\phi_0^{3(1+w_i)}
\end{equation}
the reconstructed $F(R)$ gravity of relation (\ref{r1}) in the large curvature limit, takes the following form,
\begin{equation}\label{r24}
F(R)\simeq c_1R+\frac{c_2}{\mathcal{A}}R^2+\sum_iS_i(0)\frac{\mathcal{A}^2}{R}+c_4
\end{equation}
Without great loss of generality, we can choose the coefficient of $R$ in relation (\ref{r24}) to be equal to one, that is, $c_1=1$, so by keeping leading order terms, the final form of the reconstructed $F(R)$ gravity in the large $R$ limit is the following,
\begin{equation}\label{r24c}
F(R)\simeq R+\alpha R^2+c_4
\end{equation} 
with $\alpha=\frac{c_2}{\mathcal{A}}$. This form of modified gravity is nothing else but Starobinsky $F(R)$ gravity \cite{starobinsky,sergeistarobinsky} with cosmological constant.

\subsection{$F(R)$ Gravity in the Small $R$ Limit}

In the small $R$ limit, which is actually equivalent to the large $t$ (or $\phi$ owing to the equivalence $t=\phi$ in our approximation) limit, the slow varying function $h(\phi)$ is approximately equal to $h_f$. Bearing in mind that $h_f=2/c^2$, in the large $R$ limit, relation (\ref{r15}) is written,
\begin{equation}\label{r25}
\phi\simeq \frac{2b}{3q}-\frac{2^{1/3}b^2R^{1/2}}{3}+\frac{6\times 2^{1/3q^2h_f}}{R^{1/2}}-\frac{12\times 2^{1/3}q^2h_f}{R^{1/2}}-\frac{R^{-1/2}}{3\times 2^{1/2}q^2}
\end{equation}
which at leading order becomes,
\begin{equation}\label{r25a}
\phi\simeq \left (\mathcal{A}-\mathcal{B}\right )R^{-1/2}
\end{equation}
where $\mathcal{B}$ stands for $\mathcal{B}=\frac{1}{3\times 2^{1/3}q^2}$. Thereby, the function $P(\phi )$ in the large $\phi$ limit becomes,
\begin{equation}\label{r27}
P(\phi )\simeq c_1\phi^{\frac{h_f-1+\sqrt{h_f^2+6h_f+1} }{2} }+c_2\phi^{\frac{h_f-1-\sqrt{h_f^2+6h_f+1} }{2}}+\sum_iS_i(\infty )\phi^{(-3 (1+w_i)h_f+2)\frac{1}{2}}
\end{equation}
We may write the function $P(\phi )$ in terms of the Ricci scalar by using relation (\ref{r25}). Keeping only the leading order terms for small curvatures, $P(\phi )$ becomes,
 \begin{align}\label{r28}
 P(\phi (R))\simeq c_1 \left ( \mathcal{A}-\mathcal{B}\right )^{\frac{h_f-1+\sqrt{h_f^2+6h_f+1}}{2} }R^{-\frac{h_f-1+\sqrt{h_f^2+6h_f+1}}{4}}
\end{align}
In the same vain, the function $Q(\phi (R))$ as $R$ tends to zero becomes,
\begin{align}\label{r35}
& Q(\phi (R) )\simeq  -6h_fc_1\Big{(}h_f+\frac{h_f-1+\sqrt{h_f^2+6h_f+1}}{2}\Big{)} \\ \notag &
 \times \left ( \mathcal{A}-\mathcal{B}\right )^{\frac{h_f-1+\sqrt{h_f^2+6h_f+1}}{2}-2 }R^{-\frac{h_f-1+\sqrt{h_f^2+6h_f+1}}{4}+1 }
\end{align}
Upon combining relations (\ref{r28}), (\ref{r35}) and (\ref{r1}), we obtain the reconstructed $F(R)$ gravity in the small $R$ limit,
\begin{equation}\label{r37}
F(R)\simeq -\beta R^{-\frac{h_f-1+2\delta }{4}+1 }
\end{equation}
with $\delta $ and $\beta $ being set equal to,
\begin{equation}\label{deldgsnococt}
\delta= \frac{\sqrt{h_f^2+6h_f+1}}{2},{\,}{\,}{\,}\beta=6h_fc_1\left (h_f+\delta \right ) \left ( \mathcal{A}-\mathcal{B}\right )^{\frac{h_f-3-\delta }{2}}
\end{equation}
Combining equations (\ref{r24c}) and (\ref{r37}) we obtain the following $F(R)$ gravity,
\begin{equation}\label{r39}
F(R)\simeq R+\alpha R^2+c_4+\beta R^{-\gamma}
\end{equation}
which generates the superbounce ekpyrotic Hubble parameter (\ref{hubble1}) at late times and early times. This form of the reconstructed $F(R)$ gravity is certainly of interest, since this type of $F(R)$ functions is known, under certain circumstances (see for example \cite{reviews1,importantpapers3}), to produce late time and early time acceleration. In the next section we shall investigate the late time and early time behavior of (\ref{r39}), focusing mainly on the late time, since the $R^2$ is known to produce early time acceleration.

\subsection{Early Time and Late Time Behavior of Super-bounce Generating $F(R)$ Gravity}

Having at hand the $F(R)$ gravity, given in relation (\ref{r39}), which generates the superbounce Hubble parameter in the limiting cases of early and late time, it is of interest to see the behavior of the $F(R)$ functions at late and early times. Particularly we are interested to see if inflation and late time acceleration can be achieved with this $F(R)$ gravity, working in the Jordan frame. The $R^2$ case is easy to tackle with, since this kind of gravity is a version of the Starobinsky inflation model \cite{starobinsky}, but the late time $F(R)$ has to be studied with caution, since $h_f=2/c^2$ and $c$ is restricted to take values $c>\sqrt{6}$ \cite{bounce4}, and we therefore focus on this case. Consider a late time instant, which we denote $t_{lat}$, at which $h(t)$ reaches it's limiting value $h_f=2/c^2$ value. The behavior of the total matter energy density and pressure when $t\rightarrow t_{lat}$ are approximately equal to \cite{importantpapers3}, 
\begin{align}\label{irembereveryth}
&\rho \sim \beta \Big{(}6(\gamma+1)(2\gamma +1)h_f+6(\gamma-2)h_f^2 \Big{)}\Big{(}-6h_f+12h_f^2\Big{)}^{-\gamma -1} t^{2\gamma} \\ \notag &
p=\beta \Big{(}-4\gamma (\gamma+1)(2\gamma+1)-2(8n^2+5n+3)h_f-6(n-2)h_f^2\Big{)}\Big{(}-6h_f+12h_f^2\Big{)}^{-\gamma -1}t^{2\gamma}
\end{align}
and the effective equation of state cosmological parameter $w_{eff}$ reads,
\begin{align}\label{alifeffeqn}
w_{eff}=\frac{\Big{(}6(\gamma+1)(2\gamma +1)h_f+6(\gamma-2)h_f^2 \Big{)}}{\Big{(}-4\gamma (\gamma+1)(2\gamma+1)-2(8n^2+5n+3)h_f-6(n-2)h_f^2\Big{)}}
\end{align} 
where we have set $\gamma =-\left (-\frac{h_f-1+2\delta }{4}+1 \right )$. From this we understand that acceleration can occur if $h_f>1$, which excludes the superbounce value of $h_f$, because $c>\sqrt{6}$. Therefore the late time acceleration is not generated by the term $\sim R^{-\gamma}$, at least when the superbounce generating constraint $c>\sqrt{6}$ is taken into account. Obviously if someone looses up this constraint, we can achieve late time acceleration and also, owing to the fact that the $R^2$ term is not affected by this constraint, early time acceleration can be achieved too.

\subsection{On Curvature Collineations and a Specific Super-bounce}

In the previous section we showed that when $h_f>1$, the $F(R)$ gravity that generates the superbounce solution at late times, also causes late time acceleration in the Jordan frame. In this section we also discuss a related case to this with a unique mathematical property. In order to tackle with this problem, we will need the definition of a curvature collineation. Let $M$ be a smooth spacetime with a smooth associated curvature tensor $\tilde{R}$ and $X$ a global $C^1$ differentiable vector field \cite{hull}. Let the $C^1$ diffeomorphism associated with $X$ be denoted by $\phi_t$, then $X$ is called a curvature collineation of the smooth manifold $M$, if $\mathcal{L}_XR^{a}_{bcd}=0$ and $\phi_t^*=\tilde{R}$, $\forall$ $\phi_t$ associated with $X$ \cite{hull}. Note that with $\mathcal{L}_X$ we denote the Lie dragging of the vector field $X$. Now the case in which $h_f=2$, corresponds to the metric,
\begin{equation}\label{metricformfrwhjkhcurvcoll}
\mathrm{d}s^2=-\mathrm{d}t^2+(t-b)^2\sum_i\mathrm{d}x_i^2
\end{equation}
It is obvious that if we disregard the constraint $c>\sqrt{6}$, the metric (\ref{metricformfrwhjkhcurvcoll}) is just a subcase of the FRW spacetime metric with a scale factor given in relation (\ref{basicsol1}). The Hubble parameter for the FRW metric (\ref{metricformfrwhjkhcurvcoll}) reads,
\begin{equation}\label{hubble1coll}
H(t)=\frac{2}{(t-b)}
\end{equation}
so it is similar to the Hubble parameter given in (\ref{hubble1}). Therefore, the whole argument of research we adopted in the previous sections, with regards the reconstruction of the $F(R)$ gravity, holds true for this case too, and the $F(R)$ function at late and early times is given by (\ref{r39}). However, according to the findings of the previous section, in this case the $F(R)$ function can describe early time and late time acceleration. Coming back to the collineation issue, the spacetime (\ref{metricformfrwhjkhcurvcoll}) is the only FRW spacetime which is of holonomy type $R_{15}$, with curvature rank three and also it is the only FRW spacetime which admits proper members of the curvature collineation group $CC(M)$. Actually, this group for the metric (\ref{metricformfrwhjkhcurvcoll}) consists of timelike vector fields of the form \cite{hull},
\begin{equation}\label{colvectrf}
X=f(t)\frac{\partial}{\partial t}+Z
\end{equation}
with $f(t)$ an arbitrary function $C^{\infty}$ and $Z$ any member of the 6-dimensional Killing algebra. It worths mentioning that another class of spacetimes that admits proper members of the $CC(M)$ group contains static Einstein metrics, but no other FRW metric.

\section{Ekpyrotic Loop Quantum Cosmology from $F(R)$ Gravity}

The ekpyrotic universe scenario \cite{ekpyr1,ekpyr2,ekpyr3} describes an alternative to the inflationary scenario, with scale invariant perturbations generated at a cosmic time before the Big-Bang phase. A refined scenario of the ekpyrotic cosmology was provided in \cite{ekpyr3}, where the gravitational dynamics is modified due to quantum gravity effects, in the context of LQC \cite{LQC1,LQC2,LQC3}. For informative reviews and important articles with regards to LQC see \cite{LQC1}. In the LQC context, the singularity at the Big-Bang phase is replaced by a bounce. There are various bounce models, for example see \cite{LQC2,LQC3} and for a reconstruction of bounces, cyclic cosmologies and matter bounce cosmologies from $F(R)$ theories see \cite{mbs,mbm}. Also for an account on ekpyrotic cosmology in the context of $F(R)$ gravities, see \cite{sekpd}. The LQC modified ekpyrotic scenario is realized by using the following potential for the scalar field $\phi$ \cite{ekpyr3},
\begin{equation}\label{ekpypot1}
V(\phi )=-\frac{V_0e^{\sqrt{16\pi G \rho}\phi}}{\left (1+\frac{3\rho V_0}{4\rho_c(1-3\rho )}e^{\sqrt{16\pi G \rho}\phi} \right )^2}
\end{equation}
with $0<\rho\ll 1$. The usual exponential ekpyrotic potential \cite{ekpyr1} agrees with the one given in relation (\ref{ekpypot1}), when the quantum gravity effects are considered to be negligible. There exists an ekpyrotic solution for the scalar field pressure $p_{s}$ and matter-energy density $\rho_s$ which are equal to,
\begin{equation}\label{matrstifmatr}
p_s=\omega\rho_s,{\,}{\,}{\,}\omega=\frac{2}{3\rho_s }-1
\end{equation}
which practically describes an ultra-stiff perfect fluid \cite{bounce5}. If we consider a flat FRW ekpyrotic universe with scalar potential (\ref{ekpypot1}) and with a scalar field equation of state described by (\ref{matrstifmatr}), the scale factor is equal to,
\begin{equation}\label{scalf}
a(t)=\left ( a_0t^2+1 \right )^{\rho/2}
\end{equation}
with $\rho_c$ being the critical density and $a_0=\frac{8\pi G\rho_c }{3\rho^2}$. The ekpyrotic LQC Hubble parameter corresponding to the scale factor (\ref{scalf}) is equal to,
\begin{equation}\label{r6}
H(t)=\frac{a_0\rho t}{a_0t^2+1}=\frac{h(t)}{t}
\end{equation}
with $h(t)$ being equal to,
\begin{equation}\label{r7}
h(t)=\frac{h_fqt^2}{1+qt^2}
\end{equation}
and in addition $h_f$, $q$ stand for,
\begin{equation}\label{r7a}
h_f=\rho,{\,}{\,}{\,}q=a_0
\end{equation}
The function $h(t)$ is slowly varying as it can be easily shown that it satisfies relation (\ref{r10}). We therefore apply the technique of the previous section, in order to find the $F(R)$ gravity that reproduces the Hubble parameter (\ref{r6}). In the case at hand, equation (\ref{r13}) is equivalent to,
\begin{equation}\label{r14ek}
q^2Ru^3+\left ( 2qR-6q^2+h_fq^2-2h_f^2q^2\right )u^2+\left ( R-12q+h_fq \right )u-6=0 
\end{equation}
so our aim is to solve equation (\ref{r14ek}) with respect to $u=\phi^2$. The real solution of equation (\ref{r14ek}) is,
\begin{align}\label{r15ek}
& u = \phi^2 =-\frac{2}{3 q}+\frac{2}{R}-\frac{h_f}{3 R}+\frac{2 h_f^2}{3 R}\\ \notag &
+ \Big{(}4\times 2^{1/3} q+\frac{12\times 2^{1/3} q^2}{R}+\frac{2^{1/3} R}{3}+\frac{2^{1/3}q h_f}{3}\\ \notag & -\frac{4\times 2^{1/3} q^2 h_f}{R}- \frac{8\times 2^{1/3} q h_f^2}{3}+\frac{25\times 2^{1/3} q^2 h_f^2}{3R}-\frac{4\times 2^{1/3} q^2 h_f^3}{3R}+4\times 2^{1/3} q^2 h_f^4+\frac{1}{3 2^{1/3} q^2 R}\Big{)} \\ \notag &
\times \frac{1}{\left ( \alpha_0+\alpha_1R+\alpha_2R^2+\alpha_3R^3 +\sqrt{\beta_2R^2+\beta_3 R^3+\beta_4 R^4+\beta_5R^5} \right )^{1/3}}
\end{align}
Following the line of research of the previous section, we shall find the large and small curvature limits of the above equation and we shall find the corresponding reconstructed $F(R)$ gravities in the corresponding limits.

\subsection{Large $R$ Limit}

From relation (\ref{r15ek}), by keeping leading order terms we obtain,
\begin{align}\label{r16a}
& \phi^2\simeq -\frac{1}{3 q}+\frac{\mathcal{A}_1}{R},
\end{align}
or equivalently,
\begin{equation}\label{r17a}
\phi \sim \sqrt{\frac{R-\mathcal{A}_1}{3qR}}
\end{equation}
which is valid when $R>\mathcal{A}_1$, with $\mathcal{A}_1$ being equal to,
\begin{align}\label{r18a}
\mathcal{A}_1=\frac{4\times 2^{1/3} q+\frac{2^{1/3}q h_f}{3}- \frac{8\times 2^{1/3} q h_f^2}{3}+4\times 2^{1/3} q^2 h_f^4+(2-h_f+2h_f^2)a_3^{1/3}}{a_3^{1/3}}.
\end{align}
By taking into account the fact that $h(t)$ as $\phi$ approaches zero is equal to,
\begin{equation}\label{r19}
\lim_{t\rightarrow 0}h(t)=0.
\end{equation} 
we can promptly find the function $P(\phi (R))$ in the small $\phi$ limit (which corresponds to the large $R$ limit equivalently),
\begin{equation}\label{r20}
P(\phi )=c_1+c_2\phi^{-1}.
\end{equation}
Owing to Eq. (\ref{r17a}), $\phi^{-1}$ reads,
\begin{equation}\label{r17aevanescnewsols}
\phi^{-1} \sim \sqrt{\frac{3qR}{R-\mathcal{A}_1}}
\end{equation}
so finally, the function $P(\phi )$ becomes,
\begin{equation}\label{r20newexpr}
P(\phi )=c_1+c_2\sqrt{\frac{3qR}{R-\mathcal{A}_1}}.
\end{equation}
In addition, the function $Q(\phi (R))$ in the limit where Eq. (\ref{r19}) is valid, is approximately equal to,
\begin{equation}\label{r23}
Q(\phi(R))\simeq 0.
\end{equation}
So the combination of  Eqs. (\ref{r20newexpr}) and (\ref{r23}), gives the resulting reconstructed $F(R)$ gravity of Eq. (\ref{r1}), which is,
\begin{equation}\label{r24newexpres}
F(R)\simeq c_1 R+c_2\sqrt{\frac{3qR^3}{R-\mathcal{A}_1}}
\end{equation}
The above $F(R)$ gravity can be further simplified, by expanding the resulting expression in powers of large $R$,
\begin{equation}\label{r24}
F(R)\simeq \left ( c_1+c_2\sqrt{3q}\right ) R+\frac{c_2\sqrt{3q}\mathcal{A}_1}{2}.
\end{equation}
By appropriately choosing the coefficient of the term linear to the scalar curvature as follows,
\begin{equation}\label{r24a}
c_1+c_2\sqrt{3q}=1,{\,}{\,}{\,}
\end{equation}
and by setting $\Lambda$ to be equal to,
\begin{equation}\label{r24b}
\Lambda =\frac{c_2\sqrt{3q}\mathcal{A}_1}{2},
\end{equation}
we obtain the following form of the $F(R)$ gravity,
\begin{equation}\label{r24c}
F(R)\simeq R+\Lambda,
\end{equation}
which describes the standard Einstein-Hilbert gravity with cosmological constant. Interestingly enough, this gravity describes accelerating cosmology, just as the large $R$ modified gravity description of the superbounce that we gave in the previous section.

\subsection{Small $R$ Limit}

We now turn our focus in the small $R$ limit of our approximate method, in which case the small $t$ limit of the slowly varying function $h(t)$ is equal to $h_f=\rho$. In the large $R$ limit, relation (\ref{r13}) yields,
\begin{equation}\label{r25ek}
\phi^2\simeq \frac{\mathcal{A}_2}{R}
\end{equation}
where $\mathcal{A}_2$ stands for,
\begin{equation}\label{r26ek}
\mathcal{A}_2=\left ( 25\times 2^{1/3} q^2 h_f^3+ 12\times 2^{1/3} q^2 + \frac{1}{3\times 2^{1/3} q^2 }- 4\times 2^{1/3} q^2 h_f-\frac{4\times 2^{1/3} q^2 h_f^2}{3}   \right )\frac{1}{a_0^{1/3}}
\end{equation}
Hence in this case the function $P(\phi (R))$ reads,
\begin{equation}\label{r29ek}
P(\phi )\simeq c_1 R^{-\delta}
\end{equation}
with $\delta $ being equal to,
\begin{equation}\label{r30ek}
\delta =\frac{h_f-1+\sqrt{h_f^2+6h_f+1}}{2}=1
\end{equation}
Accordingly, at leading order in the small $R$ limit, the function $Q(\phi (R))$ is equal to,
\begin{equation}\label{r36ek}
Q(\phi (R))=c_3R^{-\delta }
\end{equation}
So finally by combining relations (\ref{r29ek}), (\ref{r36ek}) and substituting to equation (\ref{r1}), the reconstructed $F(R)$ gravity in the small $R$ limit is,
\begin{equation}\label{r37ek}
F(R)\simeq c_1-c_3\frac{1}{R^{\delta }}
\end{equation}
Recalling the value of $\delta$ from (\ref{r30ek}), the most dominant term for the $F(R)$ gravity is the second one, so finally, the small $R$ reconstructed $F(R)$ gravity reads,
\begin{equation}\label{r38ek}
F(R)\simeq -c_3\frac{1}{R}
\end{equation} 
Hence by combining the resulting expressions of relations (\ref{r24cek}) and (\ref{r38ek}) we get an $F(R)$ gravity of the form,
\begin{equation}\label{r39ek}
F(R)\simeq R+\alpha R^2-c_3\frac{1}{R^{\delta }}
\end{equation}
Applying the techniques we applied in the previous section, when $\rho >1$, the function (\ref{r39ek}) can describe late time acceleration however, for the LQC ekpyrotic scenarios, $\rho\ll 1$ and therefore no late time acceleration can be consistently described too, like in the previous section's result describing the superbounce case.

\section{Conclusions}

In this paper we investigated which $F(R)$ gravities can generate two types of cosmological evolution, namely the superbounce and the LQC ekpyrotic scenario. We used a reconstruction technique which assumes that the form of the Hubble parameter can be approximated in a way so that $H(t)\sim h(t)/t$, with $h(t)$ a slowly varying function of the cosmological time. Both the cosmological models we studied lead to a Hubble parameter which is exactly of that form, so the reconstruction technique applies perfectly. Since the technique is an approximate method, we investigated the problem in the large and small curvature limits and we tried to see which $F(R)$ gravity reproduces the Hubble parameter of each cosmological scenario. As we demonstrated, the resulting picture is quite appealing since in the case of the superbounce cosmological scenario, the large curvature limit leads to a modified gravity of the form $F(R)\sim R+aR^2+c$, with $a,$ model dependent parameter. This modified gravity model is similar to the Starobinsky model \cite{starobinsky} of inflation. The small curvature limit leads to a modified gravity of the form $R^{-\delta}$, which as was explicitly demonstrated does not lead to any late time acceleration, due to the constraints put from the superbounce solution. In the case of the LQC ekpyrotic scenario, when the large curvature approximation is considered, the resulting picture is intriguingly interesting because the modified gravity is of the form  $F(R)\sim R+\Lambda$, which is the standard Einstein-Hilbert gravity description. The small curvature limit leads to an $F(R)$ gravity of the form $R^{-\delta '}$, which however again does not lead to any late time acceleration, owing to the constraints put by the LQC ekpyrosis.

In addition, for the class of cosmological theories which are described by a FRW space time with scale factor $a(t)=(at+b)^2$, which is the class where the superbounce model belongs, we showed that the spacetime has a collineation group which has only one element. This is the only non-static spacetime with this property.

Since our method heavily relies on the reproduction of the Hubble parameter and is an approximation method, it is worth examining whether similar results can be obtained if someone uses different reconstruction techniques \cite{recontechniques}. This will be done in a future work.

\section*{APPENDIX A}

In this appendix we provide the full details for the polynomial coefficients $\alpha_i,\beta_j$, with $i=1,..3$ and $j=3,...5$, appearing in relation (\ref{r15}). Recall that the $\alpha_i$'s are coefficients of the polynomial:
\begin{equation}\label{a1}
\mathcal{P}_1=\alpha_1R^2+\alpha_2R^3
\end{equation} 
while the $\beta_j$'s are coefficients of the polynomial,
\begin{equation}\label{a2}
\mathcal{P}_2= \beta_3R^3+\beta_4 R^4+\beta_5 R^5
\end{equation}
and these are given below,
\begin{align}\label{a3}
&\alpha_1=-54 b q^5 h_f-216 b q^5 h_f^2 \\ \notag & 
\alpha_2= 2 b^3 q^3 \\ \notag &
\beta_3= 23328 q^{12} h_f^3-139968 q^{12} h_f^4+279936 q^{12} h_f^5-186624 q^{12} h_f^6 \\ \notag &
\beta_4= -972 b^2 q^{10} h_f^2+38880 b^2 q^{10} h_f^3+31104 b^2 q^{10} h_f^4 \\ \notag &
\beta_5 = -1296 b^4 q^8 h_f^2 
\end{align}

\section*{APPENDIX B}

In this appendix we provide the full details for the polynomial coefficients $\alpha_i,\beta_j$, with $i=0,..3$ and $j=2,...5$, appearing in relation (\ref{r15}) Recall that the $\alpha_i$'s are coefficients of the polynomial:
\begin{equation}\label{a1}
\rho (R)=\alpha_0+\alpha_1R+\alpha_2R^2+\alpha_3R^2
\end{equation} 
while the $\beta_j$'s are coefficients of the polynomial,
\begin{equation}\label{a2}
P(R)= \beta_2R^2+\beta_3 R^3+\beta_4 R^4+\beta_5R^5
\end{equation}
and these are given below,
\begin{align}\label{a3}
& \alpha_0=432 q^6-216 q^6 h_f+468 q^6 h_f^2-146 q^6 h_f^3+156 q^6 h_f^4-24 q^6 h_f^5+16 q^6 h_f^6 \\ \notag &
\alpha_1=216 q^5-18 q^5 h_f-75 q^5 h_f^2+30 q^5 h_f^3-48 q^5 h_f^4 \\ \notag & 
\alpha_2= 36 q^4+3 q^4 h_f+30 q^4 h_f^2 \\ \notag &
\alpha_3= 2 q^3 \\ \notag &
\beta_2= 45684 q^{10} h_f^2-15228 q^{10} h_f^3+31725 q^{10} h_f^4-5076 q^{10} h_f^5+5076 q^{10} h_f^6 \\ \notag &
\beta_3 = 23004 q^9 h_f^2+594 q^9 h_f^3-12528 q^9 h_f^4-216 q^9 h_f^5 \\ \notag &
\beta_4 =3861 q^8 h_f^2+540 q^8 h_f^3-108 q^8 h_f^4 \\ \notag &
\beta_5 = 216 q^7 h_f^2
\end{align}


\begin{thebibliography}{}

\bibitem{LQC1}A. Ashtekar, P. Singh, Class. Quant. Grav. 28, 213001 (2011) [arXiv:1108.0893 ]; A. Ashtekar, Nuovo Cim. B122 (2007) 135 [gr-qc/0702030]; M. Bojowald, Class.Quant.Grav. 26 (2009) 075020 [arXiv:0811.4129]



\bibitem{LQC2}  T. Cailleteau, A. Barrau, J. Grain, F. Vidotto, Phys.Rev. D86 (2012) 087301 [arXiv:1206.6736]; J. Quintin, Yi-Fu Cai, R. H. Brandenberger, Phys. Rev. D90 (2014) 063507 [arXiv:1406.6049] ; Yi-Fu Cai, R. Brandenberger, X. Zhang, Phys.Lett. B703 (2011) 25 [arXiv:1105.4286] ; Yi-Fu Cai, R. Brandenberger, X. Zhang, JCAP 1103 (2011) 003 [arXiv:1101.0822] 


\bibitem{LQC3} J. Amoros, J. Haro, S. D. Odintsov, Phys.Rev. D87 (2013) 104037 [arXiv:1305.2344]; T. Qiu, X. Gao, E. N. Saridakis, Phys.Rev. D88 (2013) 4, 043525 [arXiv:1303.2372];  Yi-Fu Cai, Shih-Hung Chen, J. B. Dent, S. Dutta, E. N. Saridakis, Class.Quant.Grav. 28 (2011) 215011 [arXiv:1104.4349]; J. Haro, J. Amoros, JCAP 08(2014)025 [arXiv:1403.6396 ]; Yi-Fu Cai, E. Wilson-Ewing, JCAP 1403 (2014) 026 [arXiv:1402.3009 ]; J. Amoros, J. de Haro, S. D. Odintsov, Phys.Rev. D89 (2014) 10, 104010 [arXiv:1402.3071]


\bibitem{bounce1} M. Novello, S.E.Perez Bergliaffa, Phys.Rept. 463 (2008) 127 [arXiv:0802.1634]

\bibitem{bounce2} Jean-Luc Lehners, Class.Quant.Grav. 28 (2011) 204004 [arXiv:1106.0172]; Jean-Luc Lehners, Phys.Rept. 465 (2008) 223 [arXiv:0806.1245]

\bibitem{bounce3} N. Arkani-Hamed, Hsin-Chia Cheng, M. A. Luty, S. Mukohyama, JHEP 0405 (2004) 074 [hep-th/0312099]; A. Nicolis, R. Rattazzi, E. Trincherini, Phys.Rev. D79 (2009) 064036 [arXiv:0811.2197]; C. Deffayet , G. Esposito-Farese (Paris, Inst. Astrophys.) , A. Vikman, Phys.Rev. D79 (2009) 084003 [arXiv:0901.1314]; J. Khoury, B. A. Ovrut, J. Stokes, JHEP 1208 (2012) 015 [arXiv:1203.4562]


\bibitem{bounce4} M. Koehn, Jean-Luc Lehners, B. A. Ovrut, Phys.Rev. D90 (2014) 025005 [arXiv:1310.7577] 

\bibitem{bounce5} Yi-Fu Cai, D. A. Easson, R. Brandenberger, JCAP 1208 (2012) 020 [arXiv:1206.2382] 
 
 

\bibitem{ekpyr1} J. Khoury, B. A. Ovrut , N. Seiberg, P. J. Steinhardt, N. Turok, Phys.Rev. D65 (2002) 086007 [hep-th/0108187]; J. K. Erickson, D. H. Wesley, P. J. Steinhardt, N. Turok, Phys.Rev. D69 (2004) 063514 [hep-th/0312009]; Yi-Fu Cai, Shih-Hung Chen, J. B. Dent, S. Dutta, E. N. Saridakis, Class. Quantum Grav. 28 (2011) 215011 [arXiv:1104.4349] 

\bibitem{ekpyr2} Jean-Luc Lehners, P. J. Steinhardt, Phys.Rev. D87 (2013) 123533 [arXiv:1304.3122]; J. Khoury, B. A. Ovrut, P. J. Steinhardt, N. Turok, Phys.Rev. D66 (2002) 046005 [hep-th/0109050]

\bibitem{ekpyr3} E. Wilson-Ewing, JCAP 1303 (2013) 026 [arXiv:1211.6269]



\bibitem{oriti} S. Gielen, D. Oriti, L. Sindoni, Phys.Rev.Lett. 111 (2013) 3, 031301 [arXiv:1303.3576]; D. Oriti, J. P. Ryan, J. Thuerigen, arXiv:1409.3150; R. C. Helling, arXiv:0912.3011

\bibitem{piao} Y.~-S.~Piao, B.~Feng and X.~-m.~Zhang,  Phys.\ Rev.\ D bf 69, 103520 (2004) [hep-th/0310206]; Z.~-G.~Liu, Z.~-K.~Guo and Y.~-S.~Piao,  Phys.\ Rev.\ D 88, 063539 (2013) [arXiv:1304.6527]


\bibitem{vikman2} I. Sawicki, A. Vikman, Phys.Rev. D87 (2013) 067301 [arXiv:1209.2961]

\bibitem{vikman1} D. A. Easson, I. Sawicki, A. Vikman 
JCAP 1111 (2011) 021 [arXiv:1109.1047]; C. Deffayet, O. Pujolas, I. Sawicki, A. Vikman, JCAP 1010 (2010) 026 [arXiv:1008.0048]


\bibitem{importantpapers3} S. Nojiri, S. D. Odintsov, Phys.Rev. D74 (2006) 086005 [hep-th/0608008]

\bibitem{reviews1} S. Nojiri, S. D. Odintsov, Int.J.Geom.Meth.Mod.Phys. 11 (2014) 1460006 [arXiv:1306.4426]; Int. J. Geom. Meth. Mod.Phys. 4 (2007) 115 [hep-th/0601213]; S. Capozziello, V. Faraoni, Beyond Einstein Gravity, Springer, Berlin 2010;  S. Nojiri, S. D. Odintsov,  Phys.Rept. 505 (2011) 59 [arXiv:1011.0544];  S. Capozziello, M. De Laurentis, Phys.Rept. 509 (2011) 167 [arXiv:1108.6266]; K. Bamba, S. Capozziello, S. Nojiri, S. D. Odintsov, Astrophys.Space Sci. 342 (2012) 155 [arXiv:1205.3421]


\bibitem{importantpapers1} S. Capozziello, S. Nojiri, S.D. Odintsov, A. Troisi, Phys.Lett. B639 (2006) 135 [astro-ph/0604431];W. Hu, I. Sawicki, Phys.Rev.D76 (2007) 064004 [arXiv:0705.1158] ; S. Nojiri, S. D. Odintsov, Phys.Rev. D77 (2008) 026007 [arXiv:0710.1738];  S. Capozziello, V.F. Cardone, S. Carloni, A. Troisi, Int.J.Mod.Phys. D12 (2003) 1969 [astro-ph/0307018]; 


\bibitem{importantpapers2} S. M. Carroll, V. Duvvuri, M. Trodden, M. S. Turner, Phys.Rev. D70 (2004) 043528 [astro-ph/0306438]; S. Capozziello, Int.J.Mod.Phys.D11, 483 (2002) [gr-qc/0201033];  R. Myrzakulov, L. Sebastiani, S. Zerbini, Int.J.Mod.Phys. D22 (2013) 1330017 [arXiv:1302.4646]; A. Capolupo, S. Capozziello, G. Vitiello, Int.J.Mod.Phys. A23 (2008) 4979 [arXiv:0705.0319];  P. K.S. Dunsby, E. Elizalde, R. Goswami, S. Odintsov, D. S. Gomez, Phys.Rev. D82 (2010) 023519 [arXiv:1005.2205]  



\bibitem{importantpapers4}  S. Capozziello, V. F. Cardone, A. Troisi, Phys.Rev. D71 (2005) 043503 [astro-ph/0501426]; V. Faraoni, Phys.Rev. D74 (2006) 104017 [astro-ph/0610734];  S. A. Appleby, R. A. Battye, Phys.Lett.B654 (2007) 7 [arXiv:0705.3199]; S. A. Appleby, R. A. Battye, JCAP 0805 (2008) 019 [arXiv:0803.1081]; V. Faraoni,  Phys.Rev. D75 (2007) 067302 [gr-qc/0703044]

\bibitem{recontechniques} S. Nojiri, S. D. Odintsov, D. Saez-Gomez, Phys.Lett. B681 (2009) 74 [arXiv:0908.1269] ; S. Carloni, R. Goswami, P. K.S. Dunsby, Class.Quant.Grav. 29 (2012) 135012 [arXiv:1005.1840]


\bibitem{sergeinojirimodel} S. Nojiri, S. D. Odintsov, Phys.Rev. D68 (2003) 123512 [hep-th/0307288]

\bibitem{sergeibabmba} K. Bamba, S. Nojiri, S. D. Odintsov, JCAP 0810 (2008) 045 [arXiv:0807.2575]

\bibitem{chameleon} J.~Khoury and A.~Weltman,  Phys.\ Rev.\ D 69 (2004) 044026  [astro-ph/0309411];  J.~Khoury and A.~Weltman,  Phys.\ Rev.\ Lett.\   93 (2004) 171104 [astro-ph/0309300].


\bibitem{capo} Yi-Fu Cai, E. N. Saridakis, M. R. Setare, Jun-Qing Xia, Phys.Rept. 493 (2010) 1 [ arXiv:0909.2776]

 \bibitem{capo1} T. Padmanabhan, Phys.Rept. 380 (2003) 235 [hep-th/0212290]

\bibitem{peebles} P.J.E. Peebles, Bharat Ratra, Rev.Mod.Phys. 75 (2003) 559 [astro-ph/0207347]; M. Li, Xiao-Dong Li, S. Wang, Yi Wang, Commun.Theor.Phys. 56 (2011) 525 [arXiv:1103.5870]

\bibitem{faraonquin} V. Faraoni, Int.J.Mod.Phys. D11 (2002) 471 [astro-ph/0110067]; V.K. Onemli, R.P. Woodard, Class.Quant.Grav. 19 (2002) 4607 [gr-qc/0204065] 

\bibitem{tsujiintjd} S. Basilakos, S. Nesseris, L. Perivolaropoulos, Phys.Rev. D87 (2013) 12, 123529

\bibitem{wetterich} C. Wetterich, Nucl.Phys. B302 (1988) 668; C. Wetterich, arXiv:1402.5031;  C. Wetterich, Phys.Dark Univ. 2 (2013) 184 [arXiv:1303.6878]; C. Wetterich, Phys.Rev. D89 (2014) 024005 [arXiv:1308.1019]; C. Wetterich, [arXiv:1408.0156]

\bibitem{riess} A.G. Riess et al. (High-z Supernova Search Team), Astronom. J. 116, 1009 (1998) [arXiv:astro-ph/9805201]

\bibitem{planck} P.A.R. Ade et al. [arXiv:1302.5082]

\bibitem{bicep} P. A. R. Ade et al. [arXiv:1403.3985], (2014)

\bibitem{kinezosvergados} Yeuk-Kwan E. Cheung, J.D. Vergados, arXiv:1410.5710

\bibitem{oikonomouvergados}V.K. Oikonomou, J.D. Vergados, Ch.C. Moustakidis, Nucl.Phys. B773 (2007) 19 [hep-ph/0612293]; J.D. Vergados, Ch.C. Moustakidis, V. Oikonomou, AIP Conf.Proc. 878 (2006) 138-144 

 
\bibitem{starobinsky} A. A. Starobinsky, Phys.Lett. B91 (1980) 99

\bibitem{sergeistarobinsky} L. Sebastiani, G. Cognola, R. Myrzakulov, S.D. Odintsov, S. Zerbini, Phys. Rev. D 89, 023518 (2014) [arXiv:1311.0744]


\bibitem{hull} G S Hall, Symmetries and Curvature Structure in General Relativity, World Scientific Lecture Notes in Physics: Volume 46, Singapore, (2004)


\bibitem{mbs}  K. Bamba, A. N. Makarenko, A. N. Myagky, S. Nojiri, S. D. Odintsov, JCAP01(2014)008  [arXiv:1309.3748]

\bibitem{mbm} S.D. Odintsov, V.K. Oikonomou, arXiv:1410.8183

\bibitem{sekpd} S. Nojiri, S.D. Odintsov, D. Saez-Gomez, AIP Conf.Proc. 1458 (2011) 207 










\end{thebibliography}
\end{document}